\documentclass[floatfix,lengthcheck,showpacs,amssymb,amsmath,amsfonts,twocolumn,nofootinbib,longbibliography]{revtex4-1}
\usepackage{amsfonts,amsmath,units,wasysym,epsfig,graphicx,verbatim,color,subfigure,graphicx}
\usepackage{amsmath}
\usepackage{latexsym}
\usepackage{amssymb}
\usepackage{amsfonts}
\usepackage{mathtools}
\usepackage{bm}
\usepackage{color}
\usepackage{float}
\usepackage{tikz}
\usepackage{adjustbox}
\usepackage{array}
\usepackage{soul}
\usepackage{appendix}
\usepackage{physics}
\usepackage{braket}
\usepackage{xcolor}
\usepackage[normalem]{ulem}
\usepackage[none]{hyphenat}
\usepackage{lineno}

\begin{document}

\author{Sam Insley} 
\affiliation{Institute of Cosmology and Gravitation, University of Portsmouth, Portsmouth, PO1 3FX, U.K.}
\author{Michael J. Williams}
\affiliation{Institute of Cosmology and Gravitation, University of Portsmouth, Portsmouth, PO1 3FX, U.K.}
\author{Rahul Dhurkunde}
\affiliation{Institute of Cosmology and Gravitation, University of Portsmouth, Portsmouth, PO1 3FX, U.K.}
\author{Ian Harry}
\affiliation{Institute of Cosmology and Gravitation, University of Portsmouth, Portsmouth, PO1 3FX, U.K.}

\title{Normalizing flows for density estimation in multi-detector gravitational-wave searches}

\begin{abstract}
Identifying compact binary coalescences buried within the non-Gaussian and non-stationary data taken by large-scale gravitational-wave interferometers requires sophisticated multi-step search pipelines, such as the PyCBC analysis used extensively both within and outside the LIGO-Virgo-KAGRA collaborations. A critical task for these pipelines is determining the statistical significance of candidate events by comparing a ``ranking statistic" against a large background set. Currently, PyCBC’s ranking statistic incorporates the joint probability of the relative arrival times, phase delays and amplitude ratios of the signals seen in different detectors. These parameters are tightly constrained for physical signals but are much more broadly distributed for noise. PyCBC currently relies on precomputed binned histogram-based density estimators using Monte-Carlo simulations to obtain these probabilities. However, the storage requirements for these histograms scale prohibitively with the size of the detector network, preventing PyCBC from effectively analyzing four or more detectors. In this paper, we demonstrate that these histogram files can be replaced with normalizing flows, a machine learning approach to density estimation. Applying this method to data from the third observing run of Advanced LIGO and Virgo, we demonstrate that normalizing flows reduce storage requirements by more than three orders of magnitude. Furthermore, our approach maintains high sensitivity, with no more than a 0.05\% drop in the recovery of simulated signals at a fixed false-alarm rate. By relaxing several simplifying assumptions previously required by Monte-Carlo methods, we also achieved up to a 6.55\% increase in recovered signals for specific detector combinations. These results suggest that normalizing flows provide a scalable, flexible framework for the PyCBC pipeline as it expands to include four or more detectors, or to extend to searches for precessing or higher-mode signals, in future observing runs.

\end{abstract}
 \maketitle

\section{Introduction}

The field of gravitational-wave astronomy has advanced significantly since the first detection of the binary black hole merger GW150914~\cite{LIGOScientific:2016aoc} in 2015 by the Advanced LIGO detectors~\cite{LIGOScientific:2014pky}. In the decade following, the global network of detectors responsible for these observations has expanded to include Advanced Virgo~\cite{VIRGO:2014yos} and KAGRA~\cite{KAGRA:2020tym}. With plans for the construction of a fifth detector in LIGO India~\cite{Unnikrishnan:2013qwa, Iyer:2011}, this global expansion and improvement in detector sensitivity will lead to an ever-increasing rate of detections~\cite{KAGRA:2013rdx}. As of the end of the first part of the fourth observing run (O4a), 218 compact binary coalescences have been found that have a probability of astrophysical origin greater than 0.5~\cite{LIGOScientific:2025slb}. As both the detector network and number of observations grow, the likelihood of observing signals coincident in three, four or even five detectors will rise, necessitating search methods capable of simultaneous, global-network analysis.

The first step is to confidently detect these signals, which are hidden within detector data that is dominated by noise. This is possible through complex multi-step algorithms known as search pipelines~\cite{Babak:2012zx}. One such pipeline that is the focus of this paper is the PyCBC search \cite{Allen:2005fk, DalCanton:2014hxh, Nitz:2017svb, Usman:2015kfa}. PyCBC is a python-focused software package that has been used in the direct detections of hundreds of gravitational-wave signals throughout the four LIGO—Virgo—Kagra observing runs~\cite{LIGOScientific:2018mvr, LIGOScientific:2020ibl, KAGRA:2021vkt, LIGOScientific:2025yae} as well as for public data analyses such as the Open Gravitational-wave Catalogs~\cite{Nitz:2018imz, Nitz:2020oeq, Nitz:2021uxj, Nitz:2021zwj}. In order to identify signals, the PyCBC analysis applies a matched filter search~\cite{Allen:2005fk} in which a template bank of waveforms is correlated with the detector data to generate a matched-filter signal-to-noise ratio timeseries. Maxima in this signal-to-noise ratio timeseries exceeding a required threshold are identified as triggers. However, the presence of non-Gaussian and non stationary noise in the detector data can lead to non-astrophysical triggers. To reduce the contamination by noise, a $\chi^2$ test is carried out for each trigger~\cite{Allen:2004gu, Nitz:2017lco}. Furthermore, the arrival time and template parameters are subjected to a coincidence test across the detector network to ensure physical consistency~\cite{Usman:2015kfa}. Any triggers passing these tests are identified as candidate gravitational-wave events. 

 The search pipeline must then determine a measure of the statistical significance for each candidate event. Each candidate is assigned a numerical value--a ranking statistic--that denotes how likely it is to be a signal compared to noise. These values are then compared to a background of noise triggers that has been generated through time-shifting detector data~\cite{Davies:2020tsx}. This gives a measure of the number of non-astrophysical triggers with that statistic that we would expect to see in a given time window, or a false-alarm rate~\cite{LIGOScientific:2016vbw}.  Many pieces of information make up this ranking statistic~\cite{Davies:2020tsx, Kumar:2024bfe}. In this paper we focus on the term $p({\Omega}|S)$, this is the probability of the search finding a trigger with extrinsic parameters ${\Omega}$ in detectors $i \in \{1, \dots, N\}$, assuming a real signal is present. Together with the expected distribution for noise, $p({\Omega}|N)$ and various other terms~\cite{Davies:2020tsx, Kumar:2024bfe}, these give a measure of likelihood that a candidate is a signal compared to noise. A key component of $p({\Omega}|S)$ in multi-detector searches arises from the measurement of the time of arrival delay, the phase delay and amplitude ratio between detectors~\cite{Nitz:2017svb}. For a genuine astrophysical signal, we would expect the arrival times, phases and amplitudes to be highly correlated between detectors, for a background noise event, in contrast, these values are independent in each detector.
 
We do not have an analytical model for $p({\Omega}|S)$, and so instead this is estimated through the application of Monte-Carlo simulations \cite{Nitz:2017svb}. This involves sampling and storing large quantities of data so it can be later used as a look-up table. This takes the form of a N-dimensional distribution, with the number of dimensions given by $N_\text{dim}=3(N_\text{det} - 1)$, representing the relative time, phase and amplitude shifts between the $N_\text{det}$ sites. Higher dimensional distributions require larger file sizes to store the increasing amount of binned data. This becomes computationally challenging for four or more detectors, which, as we will show later, would require file sizes in the region of petabytes. As a result, PyCBC cannot currently compute the significance of coincident observations involving four or more detectors.

Here we consider an alternative method for density estimation, a type of machine learning algorithm called normalizing flows~\cite{Kobyzev:2019ydm, Papamakarios:2019fms}--a class of generative models that learn complex probability distributions by transforming a simple base density through a series of invertible, differentiable mappings. Normalizing flows have been applied to a wide range of areas in gravitational-wave science, including parameter estimation~\cite{PhysRevD.102.104057, Williams:2021qyt, Dax:2021tsq}, population studies \cite{Ruhe:2022ddi} and cosmological inference~\cite{Stachurski:2023ntw}. They are used for the modeling of complex distributions, which can then be used for both sampling and density estimation. Here we focus on the latter and demonstrate that normalizing flows can be used as an alternative to storing these large histogram files allowing us to estimate statistical significance for triggers observed simultaneously in 4 (or more) detectors.

Furthermore, we relax a number of simplifying assumptions made within the current Monte-Carlo simulations~\cite{Nitz:2017svb}. Specifically, we replace existing approximation models for time, phase, and amplitude uncertainties with more representative error distributions. We also re-evaluate our treatment of amplitude, adopting a formulation that aligns more closely with the physical response of the detector network.

This paper is structured as follows: Section \ref{sec:Sampling} gives an overview of the current sampling methodology used to estimate $p({\Omega}|S)$ and modifications are considered to improve sensitivity. Section \ref{sec:Current vs New comparison} compares the performance of this modified methodology with the current version. Section \ref{sec:Flow methodology} introduces normalizing flows as a tool for density estimation and in Section \ref{sec:Flow results} we summarise our results using the flow. In this paper, we will refer to individual detectors by representative letters as follows: LIGO Hanford (H), LIGO Livingston (L), Virgo (V) and KAGRA (K). Any combination of detectors in a network will combine these initials in the same order listed above, for example LIGO Livingston—KAGRA would be denoted as LK.

\section{Improving Sampling} \label{sec:Sampling}

In this section, we give an overview of the current methodology used to build up these multi-dimensional distributions, which closely follows the original implementation outlined by Nitz et al.~in 2017~\cite{Nitz:2017svb}. Some limitations of this methodology are discussed and some modifications are suggested to account for these.

\subsection{Current Implementation}\label{ssec:Current Implementation}

An estimation of $p({\Omega}|S)$ is made through drawing samples using a Monte-Carlo simulation.~Here, gravitational-wave signals are simulated from an isotropic distribution in sky location, polarization and inclination. The amplitude of each signal is taken to be the instantaneous maximum amplitude and is measured according to, 
\begin{equation}
\mathcal{A}_{i} = \sqrt{(\text{F}_{+\{i\}}(1+\cos^2{\iota}))^2+(\text{F}_{\times \{i\}}(2\cos{\iota}))^2},
\label{eq:amplitude}
\end{equation} 
where $\text{F}_{+\{i\}}$ and $\text{F}_{\times \{i\}}$ are the antenna response functions of a detector $i$~\cite{Thorne:1987, Forward:1978zm}, to the $\text{h}_+$ and $\text{h}_\times $ polarizations of the gravitational-wave respectively. The terms $(1+\cos^2{\iota})$ and $(2\cos{\iota})$ reflect the dependence of $\text{h}_+$ and $\text{h}_\times $ on the inclination of the source $\iota$, relative to the detector. Any common terms in the amplitude components of the gravitational-wave polarizations such as the overall distance are neglected as they will cancel out when evaluating the amplitude ratio. The amplitudes are additionally scaled by a relative sensitivity term, accounting for differences in detector PSDs. 

In addition to the amplitude, a measurement of the relative arrival time $t_i$ and phase $\phi_i$ of the gravitational-wave signal is made in each detector. For a signal at a sky location defined by its right ascension and declination, the time is measured as the delay between the signal passing through the detector and the centre of the Earth. The phase is taken to be the four-quadrant inverse tangent of the ratio of the cross to plus amplitude contributions given in Equation \ref{eq:amplitude}. At this stage no uncertainties are considered in any of the measurements made. One detector is chosen as a reference detector and then the time delay ($\Delta t$), phase delay ($\Delta\phi$) and amplitude ratio ($\frac{\mathcal{A}_1}{\mathcal{A}_2}$) are computed for every other detector relative to the reference. These values are then binned, with the bin size chosen to be proportional to the expected measurement uncertainties. A threshold is imposed, discarding any samples whose value of $\frac{\mathcal{A}_1}{\mathcal{A}_2}$ does not lie between 0.33 and 3\footnote{\label{fn:Current Values}These values are taken from those used in the PyCBC configuration for the second part of the fourth observing run (O4b).}, and the remaining samples are used to build up the distribution. 

One of the terms neglected when measuring the amplitude is source distance. To account for this, a weighting correction is applied. Each signal is weighted according to the amplitude in the detector that records the smallest signal-to-noise ratio, as this constrains the distance at which the signal will remain detectable in all detectors. Since the number of gravitational-wave signals scales with the volume of space probed by the detector network, this weighting is taken to be the cube of the amplitude.

Lastly, measurement uncertainties are incorporated through a smoothing function. This spreads out the weights in each bin according to a Gaussian kernel, with width given by the expected uncertainties in ($\Delta t$, $\Delta\phi$, $\frac{\mathcal{A}_1}{\mathcal{A}_2}$). The simulation is run and creates an individual file for each potential detector network, giving an approximation of $p({\Omega}|S)$. Each file stores a sample-set using each detector as a reference, ensuring consistent weighting and symmetries. These are then later used during gravitational-wave searches as a look-up table.

\subsection{Signal Ratio} \label{ssec:Signal Ratio}

\begin{figure}[t]
    \centering
    \includegraphics{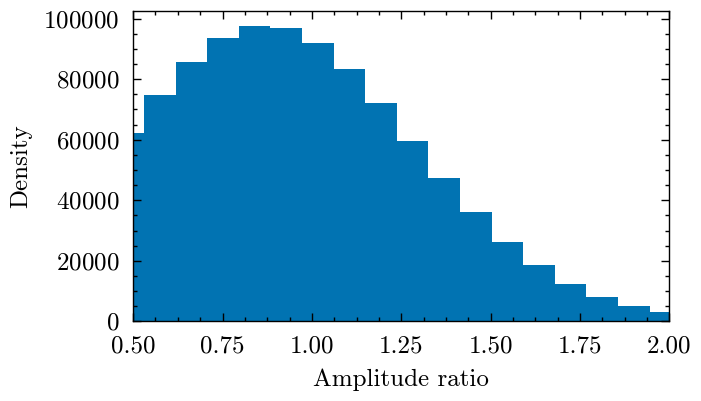}
    \hfill
    \includegraphics{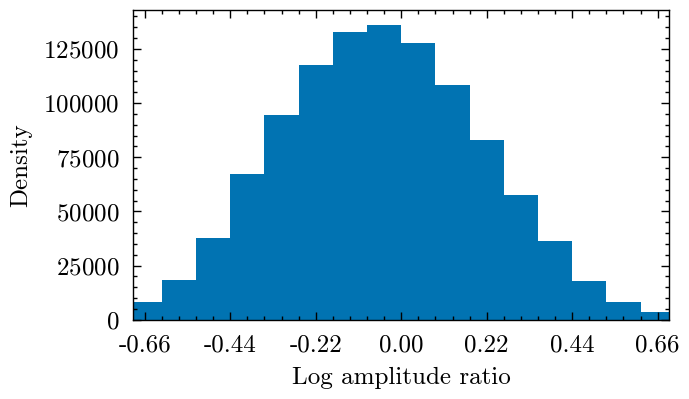}
    \caption{Amplitude-ratio sample distributions between ratios of 0.5 and 2 for LIGO Hanford—LIGO Livingston. Shown are:
    \textit{top}—linear scale;
    \textit{bottom}—log scale. }
    \label{fig:Signal ratio}
\end{figure}

In its current form, the amplitude ratio is binned uniformly on a linear scale. This can be problematic, as equal-width bins lead to an asymmetric distribution, where ratios between zero and one are allocated fewer bins than their reciprocal counterparts, as shown in Figure \ref{fig:Signal ratio} (top). This creates a bias, in which assuming a uniform population, a sample with a ratio of 0.5 will get a higher weighting than one with a ratio of 2, despite these being reciprocal. Instead we propose sampling amplitude ratio on a log scale shown in Figure \ref{fig:Signal ratio} (bottom). With this, the bins will be symmetrically distributed around an equal ratio, corresponding to zero on a log scale. 

\subsection{Distance Driven Signal-to-Noise Ratio} \label{ssec:Distance}

 Currently, the source distance of signals is not sampled. The difference in relative sensitivity between different points in the sky is included by weighting these sky points according to this sensitivity, however this does not allow us to properly include individual detector thresholds. Any sample where the amplitude ratio is less than 3:1 is discarded, but this does not properly model the behaviour of real detector networks, in particular in cases where one detector is considerably more sensitive than the other. By including distance as a parameter that we sample over we are able to place cuts on our samples that are reflective of a real search pipeline. We require a signal-to-noise ratio of five in all detectors for it to be considered "found" in that detector and a network signal-to-noise ratio of at least nine\footnote{One could of course choose different values here based on other scenarios with higher, or lower, detection thresholds}. In the case of HV, we find that the current amplitude ratio cuts would discard $\sim20\%$ of signals that are retained when thresholding with more physically realistic signal-to-noise ratio thresholds, producing a more accurate measurement of the probability distribution for this network. In addition, sampling in distance will also allow us to more accurately model the uncertainties in $(\Delta t, \Delta\phi, \mathcal{A}_1/\mathcal{A}_2)$, as outlined in Section \ref{ssec:Uncertainties}.

 In order to implement this distance-driven approach, luminosity distances $d_L$ are drawn from a power law distribution with a probability density $p(d_L)\propto d_L^2$, up to a maximum distance $D_{max}$\footnote{This distance should not be interpreted as physical distances, and does not have any mass dependence, but is simply an amplitude scaling that reflects the distribution of distance for compact binary mergers. We do not consider cosmological effects here.}. The amplitudes are then scaled according to $\mathcal{A}_i\propto\frac{1}{d_L}$. $D_{max}$ is chosen so that the tail of the distribution of amplitudes, based on the relative sensitivity of the least sensitive detector in the network, extends past the commonly used signal-to-noise ratio detection threshold of five by requiring that no more than 1\% of signals with amplitude greater than five originate from distances within 5\% of $D_{max}$. This is further described in Appendix \ref{app:L1amp}.

\begin{table}[t]
\centering
\begin{tabular}{|c|c|c|}
\hline
\textbf{Detectors} & \textbf{Modified Sampler}  & \textbf{Normalizing Flow} \\ 
\hline
HL & -0.659  & -0.997 \\
LV &  -0.875  & -0.999 \\
HV & -0.968  & -1.20 \\
HLV & - & -4.07 \\ 
\hline
\end{tabular}
\caption{Additional statistic corrections applied to coincident triggers in different detector combinations. Values are shown for the modified sampler using the original histogram-based density estimator as well as for the normalizing flow trained on the modified samples. Due to computational restrictions, the HLV file for the modified case was kept consistent with the original so no correction was needed.}
\label{tab:Statistic corrections}
\end{table}

By removing the previous amplitude-ratio bounds when sampling, the total volume of the parameter space was expanded. Since the noise distribution $p(\Omega|N)$ is assumed to be uniform, its density is inversely proportional to this volume; consequently, any change in volume shifts the likelihood ratio $p(\Omega|S)/p(\Omega|N)$ by a constant factor. To ensure consistent normalization when comparing methodologies, we apply additive `statistic corrections' to all triggers within a given detector network (see Table \ref{tab:Statistic corrections}). These corrections are determined numerically as the additional volume sampled over in signal ratio compared to the original histogram files. This aligns the background distributions across the original and modified sampling methods, including those generated via normalizing flows as we will discuss later. Further work is suggested to investigate the normalization framework, specifically the assumption that noise is distributed uniformly over the signal ratio space, to attempt to eliminate the need for these manual offsets.

\subsection{Measurement Uncertainties} \label{ssec:Uncertainties}

The presence of noise, which we assume in creating these probability distributions to be Gaussian, induces a measurement uncertainty in the arrival time, phase and amplitude in each detector. In the current PyCBC search, these measurement uncertainties are incorporated into the phase-time-amplitude sampling through the joint process of binning and smoothing. For each dimension, the uncertainty is applied through an independent Gaussian, with width equal to the expected error in that particular measurement. However, as we will show, there are known correlations in the expected uncertainties for ($t$, $\phi$, $\mathcal{A}$) measured in a given detector, which need to be accounted for. In addition, these uncertainties all depend on the signal-to-noise ratio of the signal, and so using a single Gaussian with a defined width for all signals is not ideal. Here we propose incorporating these correlated measurement uncertainties directly into the sampling of arrival time, phase and amplitude, as explained below.

\subsubsection{Uncertainty Correlations}

\begin{figure*}[t]
    \centering
    \includegraphics{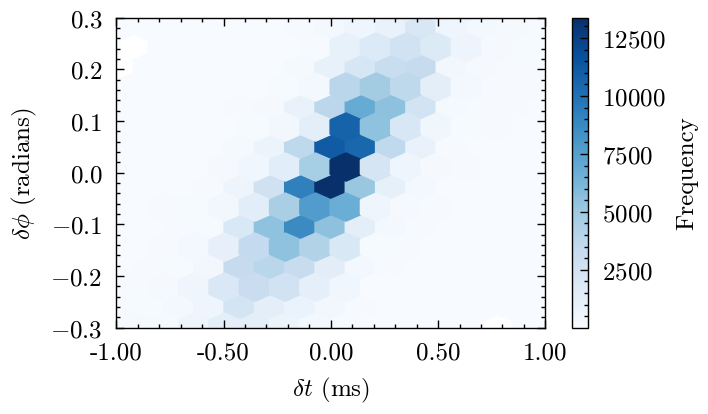}
    \hfill
    \includegraphics{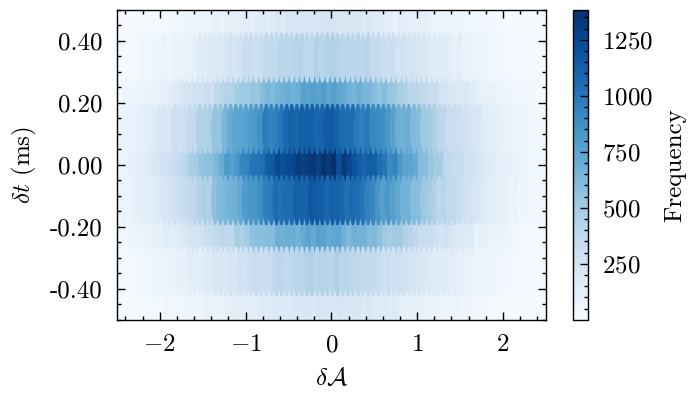}
    \hfill
    \includegraphics{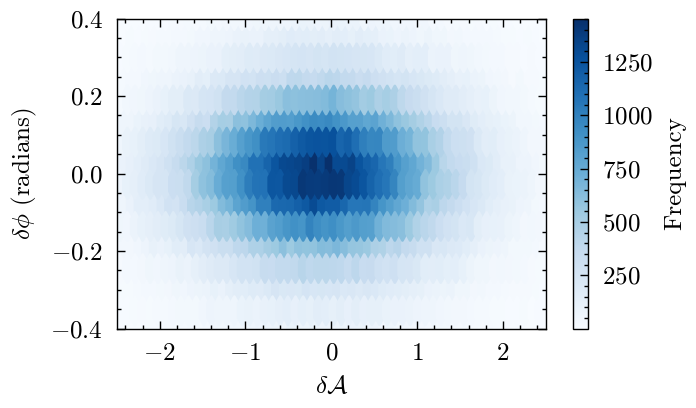}
    \caption{Distribution of the: 
    \textit{top}—time and phase,
    \textit{bottom left}—time and signal-to-noise ratio and 
    \textit{bottom right}—phase and signal-to-noise ratio uncertainties  for 300,000 simulated gravitational-wave signals with signal-to-noise ratio less than 20.}
    \label{fig:unc_correlations}
\end{figure*}

To measure the correlation between the noise-induced measurement uncertainties on $t$, $\phi$ and $\mathcal{A}$ we simulate 10,000 iterations of colored noise with a simulated signal added to the noise. The component masses and distances of the simulated signals are sampled from uniform distributions in the ranges [0,80]~$M_\odot$ and [0,5000]~Mpc respectively. We then perform matched-filtering to measure the time, phase and amplitude of the recovered signal at the peak signal-to-noise ratio. The discrepancies between the noiseless values compared to the recovered $(\delta t, \delta\phi, \delta \mathcal{A})$ is shown in Figure \ref{fig:unc_correlations}.

\begin{table}[t]
\centering
\begin{tabular}{|c|c|c|}
\hline
\textbf{Variables} & \boldmath{$r$}  & \boldmath{$p$} \\ 
\hline
$\delta t$ vs.\ $\delta \phi$ & $8.62\times 10^{-1}$  & $<10^{-16}$ \\
$\delta \mathcal{A}$ vs.\ $\delta t$ &  $-4.08\times 10^{-4}$  & $7.79\times 10^{-1}$ \\
$\delta \mathcal{A}$ vs.\ $\delta \phi$ & $-6.75\times 10^{-5}$  & $9.63\times 10^{-1}$ \\
\hline
\end{tabular}
\caption{Results of Pearson's correlation. Extremely small $p$-values are reported as bounds due to numerical precision limits.}
\label{tab:correlations}
\end{table}

We perform a Pearson's r test for correlation, the results of which are shown in Table \ref{tab:correlations}. The results show a significant correlation between $\delta t$ and $\delta\phi$, but little correlation between $\delta \mathcal{A}$ and either of the other parameters. However, it is important to note that $\delta t$ and $\delta\phi$ are dependent on the overall signal-to-noise ratio. For this work, we therefore decide to take into account the correlation between  $\delta t$ and $\delta\phi$ by drawing values of these uncertainties from a single bivariate Gaussian and independently drawing $\delta \mathcal{A}$.

\subsubsection{Phase and time uncertainty} \label{ssec:phaseandtimeunc}

To be able to draw the phase and time uncertainty from a bivariate Gaussian we need to define it's mean and covariance as given by
\begin{equation}
    \mu = \begin{pmatrix}\mu_t\\\mu_\phi\end{pmatrix}, \quad \Sigma=\begin{pmatrix} \sigma^2_t & r\sigma_t\sigma_\phi \\ r\sigma_t\sigma_\phi & \sigma^2_\phi \end{pmatrix},
\label{eq: bivariate}
\end{equation}
where ($\mu_t$, $\mu_\phi$) and ($\sigma_t$, $\sigma_\phi$) are the means and standard deviations of $\delta_t$ and $\delta_\phi$, here $r$ is the correlation coefficient given in Table \ref{tab:correlations}. Both $\mu_t$ and $\mu_\phi$ are taken to be zero as the uncertainties will be centered on the sampled values. However we need to choose appropriate values for $\sigma_t$ and $\sigma_\phi$. In 2009, Fairhurst \cite{Fairhurst:2009tc} showed that $\sigma_t$ can be well approximated as a Gaussian with width
\begin{equation}
\sigma_t = \frac{1}{2\pi\rho\sigma_f},
\label{eq:fairhurst2009}
\end{equation}
where $\sigma_f$ is the bandwidth of the signal given by,
\begin{equation}
\begin{split}
\sigma_f^2 =& \left(\frac{4}{\rho^2}  \int_{0}^{\infty} \frac{f^2 |h(f)|^2}{S(f)} \, df\right)\\ & - \left(\frac{4}{\rho^2} \int_{0}^{\infty} \frac{f |h(f)|^2}{S(f)} \, df \right)  ^2 .
\end{split}
\label{eq:bandwidth}
\end{equation}
Here, $S(f)$ is the one sided noise power spectral density and $h(f)$ denotes the frequency-domain gravitational waveform. In its current form, a measure of bandwidth is difficult to incorporate directly into the sampling process, here we measure $\sigma_t$ using Equation \ref{eq:fairhurst2009}, choosing a constant bandwidth of 30 Hz. This value was chosen so that an event with a signal-to-noise ratio of five has a $\sigma_t$ consistent with the current implementation. Further work is recommended to introduce a more direct measurement of bandwidth.

For determining $\sigma_\phi$, we adopt a numerical approach, where we look to fit a model according to $\sigma_\phi=\frac{k}{\mathcal{A}}$, where $k$ is a constant to be found. A linear regression model was fit to $\sigma_\phi$ and the inverse of $\mathcal{A}$, measured from the matched-filter search of 50 training waveform templates. From this we find $k=2.20\pm 0.040$, this is validated against the measurements from 20 independent validation waveforms and we find agreement in the fit to within 1\%. 

\subsubsection{Signal-to-noise ratio uncertainty} \label{sssec:signal-to-noise ratio_unc}

Finally, we define $\delta \mathcal{A}$. In matched-filter searches the signal-to-noise ratio is commonly expressed as
\begin{equation}
\rho(t)^2 = (s|\hat{h}_\text{+})^2 + (s|\hat{h}_{\times})^2,
\label{eq:snr}
\end{equation}
where we use the standard definition of the inner product
\begin{equation}
(a|b) = 4\text{Re}\int_0^\infty{\frac{\tilde{a}(f)\tilde{b}^*(f)}{S(f)}e^{2\pi if}\text{df}}.
\label{eq:inner product}
\end{equation}
Here $S(f)$ is the one-sided power spectral density of the detector noise and $\tilde{s}(f)$ is the Fourier transformed detector data.
$\hat{h}_\text{+}$ and $\hat{h}_{\times}$ represent the $+$ and $\times$ component of the expected gravitational-wave signal, which are assumed to be orthogonal $(\hat{h}_{\times}|\hat{h}_\text{+}) = 0$ and normalized $(\hat{h}_{\times}|\hat{h}_{\times}) = (\hat{h}_\text{+}|\hat{h}_\text{+}) = 1$.

If the detector noise is Gaussian the recovered signal-to-noise will be shifted according to noise contributions in both polarizations.
\begin{equation}
    \rho^2 = (({h}_\text{+}|\hat{h}_\text{+})+(n|\hat{h}_\text{+}))^2 + (({h}_{\times}|\hat{h}_{\times})+(n|\hat{h}_{\times}))^2,
\label{eq:noisy amp}
\end{equation}
where since $\hat{h}_\text{+}$ and $\hat{h}_{\times}$ are orthogonal, then $(n|\hat{h}_\text{+})$ and $(n|\hat{h}_{\times})$ are statistically independent Gaussians with mean zero and width one. Following this definition, we incorporate measurement uncertainties into the amplitude defined in Equation \ref{eq:amplitude}. For each simulated signal, two samples are drawn per detector from independent Gaussians with mean zero and width of one. These are then separately added to the $+$ and $\times$ dependent components in Equation \ref{eq:amplitude}. The noise corrections outlined here could similarly be used to measure $\delta\phi$, giving a good approximation if independent of $\delta t$, we choose to instead use the approach outlined in Section \ref{ssec:phaseandtimeunc}, which includes this dependency.

\section{Improvements in Detection Rate for Coincident Events} \label{sec:Current vs New comparison}

\begin{figure*}[t]
    \centering
    \includegraphics{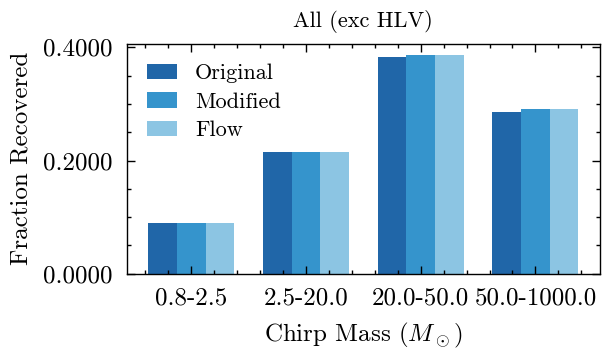}
    \hfill
    \includegraphics{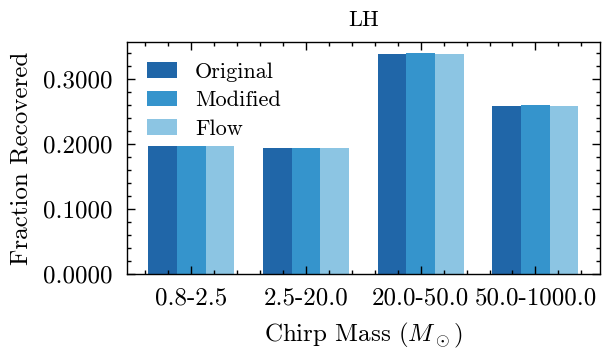}

    \vspace{0.6em}
    
    \includegraphics{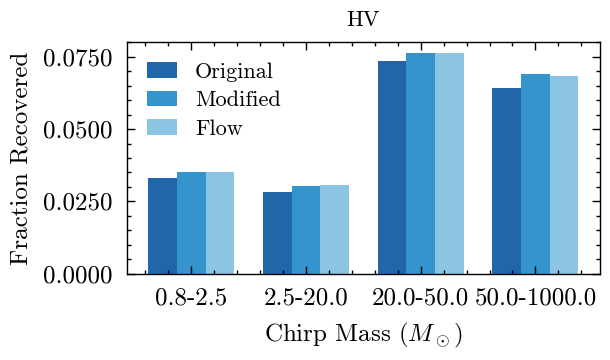}
    \hfill
    \includegraphics{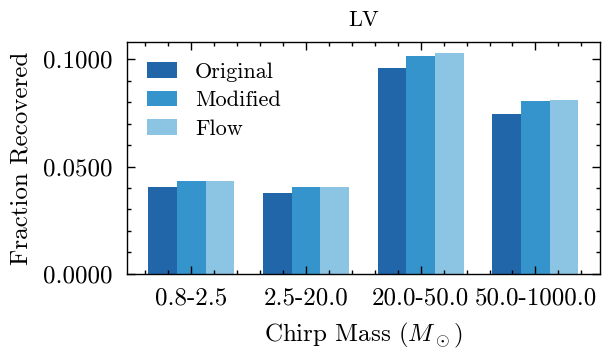}
    
    \caption{Fraction of found injections with false-alarm rate less than one per year as a function of chirp mass for different detector combinations. Each subplot shows three datasets corresponding to the original methodology, the modified sampling methodology outlined in Section \ref{sec:Sampling} using the original histogram-based density estimator and the modified sampler using a normalizing flow for density evaluation. Shown are:
    \textit{top left}—all found (excluding HLV);
    \textit{top right}—HL;
    \textit{bottom left}—HV;
    \textit{bottom right}—LV.}
    \label{fig:2DET_found_injs}
\end{figure*}

In this section, we evaluate the performance of the modifications proposed to the sampling methodology outlined in Section \ref{sec:Sampling}. This is done by measuring the sensitivity of PyCBC offline searches \cite{Usman:2015kfa} to gravitational-wave signals. Search sensitivity is estimated by recovering simulated signals, known as injections, from the real detector data. A large number of injections are added to the detector data and the search's ability to recover these is used to quantify the sensitivity \cite{Usman:2015kfa}. 

We run two offline PyCBC searches on the third observing run, utilizing publicly available data from both the Advanced LIGO and Advanced Virgo detectors \cite{KAGRA:2023pio}. Specifically, we analyse detector data taken between 7th February 2020 at 17:49:37 and 14th February 2020 at 20:30:35. These searches utilize the ranking statistic used by PyCBC searches during the second part of the fourth observing run (O4b) \cite{LIGOScientific:2025slb}. The first run uses the original phase-time-amplitude histogram files for HL, HV and LV, with the second run replacing these with ones incorporating the suggested improvements from Section \ref{sec:Sampling}. We are unable to run a HLV analysis with these changes as generating enough samples to fully populate our binned histograms is challenging, and the normalizing flow, which we demonstrate in the next section, solves this problem for us. Injections were added to the detector data, with parameters drawn from the same astrophysically-motivated distribution used to produce LVK observational constraints on compact binary rates and population models~\cite{KAGRA:2021duu}. We then compare the number of gravitational-wave injections recovered above with a false alarm rate of less than one per year to evaluate how these changes affect our sensitivity.

Figure \ref{fig:2DET_found_injs} shows the fraction of injections recovered in certain chirp mass bins\footnote{Figure \ref{fig:2DET_found_injs} includes data from the original and modified sampling methodologies but also using normalizing flows for density estimation, the latter is discussed in Section \ref{sec:Flow results}.}. Results are presented for different combinations of operating detectors, together with the total fraction of recovered injections (excluding HLV analyses). For each detector combination, the results show \emph{only} those injections recovered in each of the operating detectors. However, the total fraction of recovered injections shown in Figure \ref{fig:2DET_found_injs} (top left) includes both single and two detector coincident events. We recover an additional 19 injections, corresponding to a 0.55\% increase in the total number of found injections for the new sampling methodology. This percentage increase is dominated by HL signals, as these are the most common detections. For HL we find the rate of injections recovered is within 0.08\% for the two methodologies. However, much larger increases are recorded for HV and LV coincidences where we find increases of 6.55\% and 6.09\% respectively. These increases are observed across all four of the chirp mass bins used, with the largest increases in sensitivity occurring in the $20-50$ M$_\odot$ and $50-1000$ M$_\odot$ ranges.

\section{Normalizing Flows for Density Estimation} \label{sec:Flow methodology}

The current PyCBC search uses a histogram-based density estimator to approximate $p({\Omega}|S)$, which is part of the ranking statistic for gravitational-wave triggers. This becomes unfeasible for four or more detectors, where the dimensional scaling results in storage requirements that are not manageable within the PyCBC search. In this section, we give an overview of normalizing flows, a flexible alternative for density estimation that would replace the large binned sample files with a set of model parameters. The implementation of the normalizing flow is outlined and we discuss some of the potential challenges involved.

\subsection{Overview}\label{ssec:Flow overview}

Normalizing flows are a generative machine learning model that model complex, multi-dimensional distributions \cite{Tabak_2010,Tabak:2013cnz, Kobyzev:2019ydm,PhysRevD.102.104057} and can then be used for both density estimation and sample generation.
Starting from a simple latent distribution $p_U(u)$, such as a multivariate normal, the true data distribution $p_X(x)$ can be approximated through a sequence of transformations whose composition is denoted by $g$, where $g(U)=X$.
Evaluating the density $p_X(x)$ requires both the probability of the corresponding latent variable under the base distribution and the change in volume induced by the transformation $g$. Therefore, $g$ must be invertible and differentiable. The density can then be computed using the change-of-variables formula:
\begin{equation}
    p_X(\textit{x}) = p_U(g^{-1}(\textit{x}))\left|\text{det}\left(\frac{\partial  g^{-1}(\textit{x})}{\partial \textit{x}}\right)\right|
\label{eq:change of variables}
\end{equation}
where $\text{det}\left(\dfrac{\partial g^{-1}(\textit{x})}{\partial \textit{x}}\right)$ is the Jacobian determinant of the transformation. Each transformation is parametrized by a neural network with weights and biases $\Psi$, which are optimized by minimizing the Kullback–Leibler divergence \cite{Papamakarios:2019fms} between the target and modeled distributions.

\subsection{Implementation}\label{ssec:Flow implementation}

In this work, we use a Neural Spline Flow \cite{Durkan:2019nsq} with coupling transforms~\cite{Dinh:2015} and Rational Quadratic Splines, as implemented in the Python library \texttt{glasflow} \cite{Williams:2024glasflow} which has built upon \texttt{nflows} \cite{nflows}. Neural spline flows were chosen for their high expressiveness and their ability to support a uniform latent space. While normalizing flows are typically trained using a multivariate normal latent distribution, we instead adopt a multivariate uniform distribution because both time and phase delay are inherently bounded. This choice avoids the difficulty of mapping a Gaussian distribution with support over $\mathbb{R}^{n}$ to a target distribution defined on a finite interval \cite{Papamakarios:2019fms}.

To train the model, samples are generated through the same process described in Section \ref{ssec:Current Implementation}, incorporating the modifications proposed in Sections \ref{ssec:Signal Ratio}, \ref{ssec:Distance} and \ref{ssec:Uncertainties}. Unlike the previous approach, the samples no longer need to be binned as the normalizing flow models a continuous distribution. For the two-- and three-detector distributions, the normalizing flow is trained on 500,000 samples, with the corresponding training parameters and computational cost detailed in Table \ref{tab:flow training parameters}. During training, the neural network predicts the parameters of the rational quadratic splines (the knot positions, heights, and derivatives) that define the coupling transformations \cite{Durkan:2019nsq}. These parameters are optimized by minimizing the loss function using the Adam optimizer~\cite{Kingma:2014vow} with a learning rate of 0.001. The hyperparameters were selected through repeated testing to ensure accurate modeling of the distributions while minimizing the computational cost of density evaluation. The trained model is then used during the search to evaluate the probability density of triggers via Equation \ref{eq:change of variables}. If significant changes are observed in detector PSDs, the normalizing flow would need to be retrained on a new set of samples reflecting the new relative sensitivities between detectors. Similarly to re-generating the histogram files, we find that training the normalizing flow does not exhibit a significant cost as shown in Table \ref{tab:flow training parameters}.

\begin{table}[t]
\centering
\begin{tabular}{|c|c|c|}
\hline
\textbf{Training Parameters} & \textbf{2 Detector}  & \textbf{3 Detector} \\ 
\hline
Number of transforms & 4  & 4 \\
Number of neurons &  10  & 80 \\
Latent distribution & Uniform  & Uniform \\
Number of bins & 4 & 15 \\
Number of training samples & 500,000 & 500,000 \\
Training Cost (CPU hours) & 0.25 & 1.25 \\
\hline
\end{tabular}
\caption{Parameters used to train the normalizing flow model. This is based on the number of detectors present in the target distribution. We detail the training parameters for four and five detector cases in Appendix \ref{app:training parameter 4+}.}
\label{tab:flow training parameters}
\end{table}

\section{Normalizing Flow Performance and Computational Efficiency} \label{sec:Flow results}

In this section, we evaluate the performance of normalizing flows relative to the traditional histogram-based methodology. We assess both the effectiveness of the flow-based density estimation in modeling the phase-time-amplitude distributions and the comparative computational efficiency of the two approaches

\subsection{Performance}\label{ssec:Flow Performance}

We evaluate the performance of the normalizing flow methodology by the same measure used in Section \ref{sec:Current vs New comparison}. The statistic corrections applied during the search are shown in Table \ref{tab:Statistic corrections} and the results are shown in Figure \ref{fig:2DET_found_injs}. Since the normalizing flow was trained on the modified sampling methodology, this is our main point of comparison. We find that for two detector cases we observe only small perturbations $\lesssim 0.05\%$ in the number of recovered injections when compared to the results using the modified sampling methodology indicating that the new flow-based methodology maintains search sensitivity. The performance here is mainly influenced by the high density regions where the bulk of signals lie, in Appendix \ref{app:density comparison} we compare the densities evaluated by the two methodologies across the whole parameter space. 

\begin{figure}[t]
    \centering
    \includegraphics{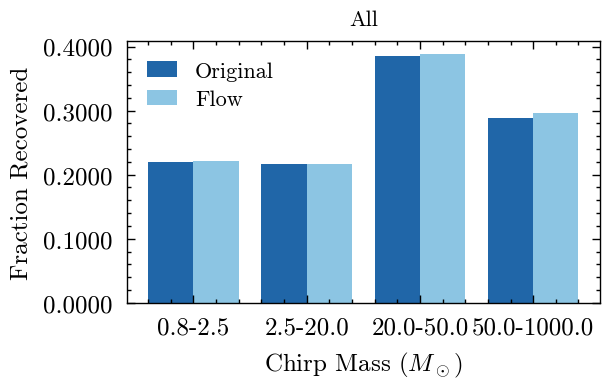}
    \hfill
    \includegraphics{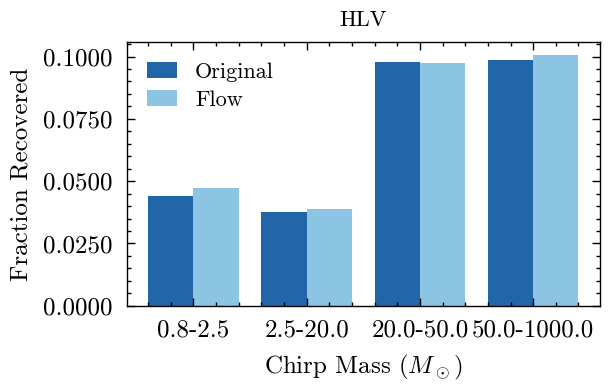}
    \caption{Fraction of found injections with false-alarm rate less than one per year as a function of chirp mass. Each subplot shows two datasets corresponding to the histogram-based density estimator using the original sampling methodology and the normalizing flow using the modified sampling methodology outlined in Section \ref{sec:Sampling}. Shown are:
    \textit{top}—all found injections;
    \textit{bottom}—injections recovered as HLV triggers.}
    \label{fig:3DET_found_injs}
\end{figure}

With the normalizing flow implemented we can also compare the performance for HLV candidates, albeit here we are comparing to the original results and so any changes observed are due to a combination of both the modified sampling as well as the normalizing flow. This is shown in Figure \ref{fig:3DET_found_injs}, which presents the total fraction of injections recovered across all detector combinations (top), together with the fraction recovered as a HLV coincident trigger (bottom). In total, we recover 35 more injections with a false alarm rate less than 1 per year, corresponding to a $0.78\%$ increase compared to the original methodology, while the number of HLV recoveries are found to increase by $3.85\%$.

\subsection{Four Detector Search}\label{ssec:4det}

One of the main motivations for this study was to be able to analyse data from four or more detectors. We evaluate the performance of the normalizing flow in this case by performing the first full, PyCBC coincident search on 4-detector data. An offline PyCBC search was conducted on simulated strain data for detectors located at LIGO Livingston, LIGO Hanford, Virgo and KAGRA. Here we use a consistent power spectral density across the detectors, assuming an equal sensitivity across the detector network. The search is run with settings consistent to those used within the fourth observing run (O4b). The injection set used was simulated from a broad parameter space of binary black holes to test the ability of the search to recover signals in this space.

Due to computational constraints, a full analysis on four or more detectors is not possible with the current histogram-based approach and therefore we do not have an existing benchmark to compare our results against. The distributions of the training data alongside the modeled distribution from the normalizing flow are included for four and five detector cases in Appendix~\ref{app:training parameter 4+}.

In Figure~\ref{fig:4detcomparison} we show the background distributions for each multi-detector combination (top) and the distribution of found and missed injections (bottom). The backgrounds show smooth, continuous distributions for all detector combinations including HLVK, indicating the normalizing flow has likewise learned a sufficiently smooth distribution in these cases. For a fixed number of detectors, the background distributions largely overlap, as expected for detectors with comparable sensitivities. Any deviations are consistent with detector observing times as well as differences in the noise response to each phase-time-amplitude distribution. The found and missed injections follow the expected relation with combined optimal SNR, with more injections recovered at a lower false-alarm rate for larger SNRs. In total, 511 injections were recovered as 4 detector coincidences, with measured false-alarm rates between 0.5 and $3.03 \times 10^{-4}$. Further work would need to be done to better optimize the PyCBC search for a four detector network, however the results from this initial test demonstrate that a full analysis of four detector data is now possible using the normalizing flow methodology.

\begin{figure}[t]
    \centering
    \includegraphics{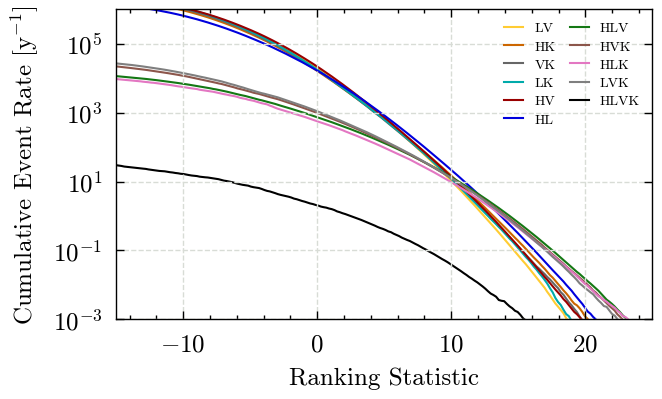}
    \hfill
    \includegraphics{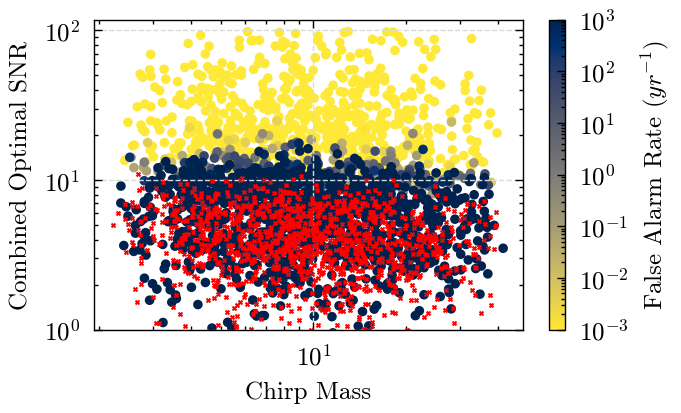}
    \caption{Results from a four detector search on simulated data. Shown are:
    \textit{top}—Cumulative rates of noise events in specified detector combinations;
    \textit{bottom}—Found and missed injections distributed according to their chirp mass, optimal signal-to-noise ratio and false-alarm rate. Missed injections are shown as red crosses. Only one third of the total injections are shown in order to increase visual clarity, these were chosen randomly.}
    \label{fig:4detcomparison}
\end{figure}

\subsection{Computational Cost and Storage}\label{ssec:Flow Cost}

\begin{figure}[t]
    \centering
    \includegraphics{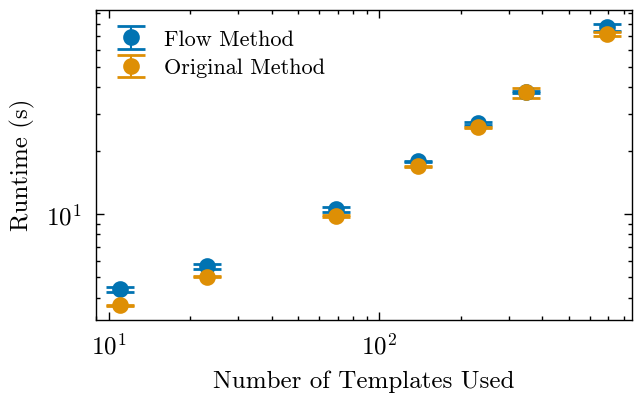}
    \hfill
    \includegraphics{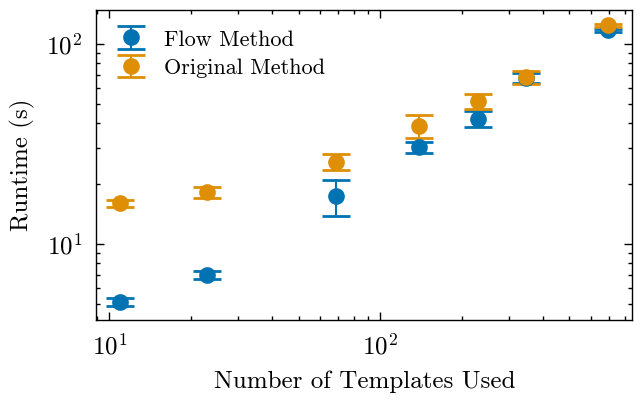}
    \caption{The mean runtime and standard deviation for the PyCBC executable \texttt{pycbc\_coinc\_findtrigs} using the original and normalizing flow methodologies as a function of the number of templates used. Shown are:
    \textit{top}—two detector search;
    \textit{bottom}—three detector search.}
    \label{fig:Runtime_cost}
\end{figure}
\begin{figure}[t]
    \centering
    \includegraphics{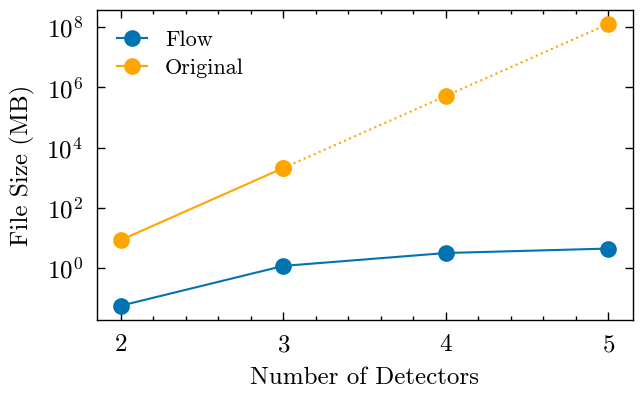}
    \caption{The file size required to store the information needed for density estimation as a function of the number of detectors in the network. Shown are the sizes needed for the original histogram-based density estimator and the normalizing flow method. The dotted line indicates an extrapolated estimate assuming the fractional increase from 3$\rightarrow$4 and 4$\rightarrow$5 detectors matches that observed between 2$\rightarrow$3 detectors.}
    \label{fig:File_size}
\end{figure}

We evaluate computational efficiency across two metrics: execution runtime and storage requirements. To measure execution runtime we look at the time taken to run the PyCBC executable \texttt{pycbc\_coinc\_findtrigs}, which reads in single detector trigger files for a given waveform template, identifies any which are coincident across detectors and then computes a chosen ranking statistic for them. These tests were all completed on the same machine, running on a single CPU without other processes running in parallel. The time taken is shown in Figure \ref{fig:Runtime_cost}, as a function of the number of waveform templates used in the search. We do not find any significant changes ($<10\%$ at large numbers of templates) to the runtime in either the two or three detector case, with the later showing a slight decrease in runtime for the normalizing flow methodology.

The most significant improvement occurs in the reduced storage requirements. Figure \ref{fig:File_size} shows the file sizes required to store the phase-time-amplitude information as a function of the number of detectors. The original methodology required a file size of 2.1 GB for a three detector case, an increase of over two orders of magnitude than the 8.6 MB needed for two detectors, extrapolating to 4 (or even 5) detectors results in O(TB) (or even O(100 TB)) file sizes. In contrast, the normalizing flow's size is determined by its parameter count, which does increase with both the dimensionality and complexity of the target distribution and so will scale with the number of detectors. However, utilizing a normalizing flow significantly reduces the amount of storage needed to 59 KB and 1.2 MB for the two and three detector cases respectively, remaining O(MB) for the 4 and 5 detector cases. Therefore this methodology presents an approach that allows us to analyse an expanding global network of detectors with the PyCBC search technique.

\section{Conclusions and Future Work}

In this work, we address the computational bottlenecks preventing the PyCBC search pipeline from scaling beyond three-detector networks. By replacing traditional histogram-based density estimators with normalizing flows, we have reduced the storage requirements for multi-detector ranking statistics by over three orders of magnitude while not compromising sensitivity. While the current histogram-based density estimators would reach the terabyte or petabyte scale for four- and five-detector networks, the flow-based approach remains manageable at $<10$ MB, effectively removing the current dimensionality constraint. We have, for the first time, demonstrated an end-to-end analysis of 4-detector data using the PyCBC search procedure.

The normalizing flow approach also allowed us to relax a number of assumptions in the Monte-Carlo sampling of the underlying PDF. By better modelling the amplitude ratio, sampling directly in distance and better including measurement uncertainties we observe an increase in sensitivity to compact binary mergers. These changes were most notable in detector combinations of unequal sensitivity, e.g. if only Hanford and Virgo, or Livingston and Virgo are operating, where up to a $\sim6\%$ increase in sensitivity was observed.

In addition to the benefits discussed here, normalizing flows will allow for the integration of more complex physics into the phase-time-amplitude distributions. The storage limitations of the histogram-based density estimators have also been highlighted in precessing searches \cite{Dhurkunde:2026jhp}. There is also the case of early-warning alerts \cite{Magee:2021xdx, Nitz:2020vym}, where the use of cut-off frequencies introduces frequency-dependent timing uncertainties that should be accounted for within the sampling. Using Normalizing Flows we could include this information through a measure of bandwidth. Finally, it would be beneficial to include information on the status of any detectors that a given trigger was not recorded in, for example the distribution of triggers \emph{only} found by Hanford and Virgo when Livingston is operating is quite different to the distribution when Livingston is operating. Normalizing flows could offer a way to incorporate this additional science without the storage issues faced by the histogram-based approach in higher dimensions.

\acknowledgements

SI thanks the STFC for providing support through the STFC-funded Doctoral Landscape PhD studentship scheme. MW, RD and IH acknowledge support through STFC's gravitational-wave grant scheme via awards ST/V005715/1, ST/Y005260/1 and UKRI2490. 
The authors are grateful for computational resources provided by the LIGO Laboratory and supported by National Science Foundation Grants PHY-0757058 and PHY-0823459. This material is based upon work supported by NSF's LIGO Laboratory which is a major facility fully funded by the National Science Foundation. For the purpose of open access, the authors have applied a Creative Commons Attribution (CC BY) licence to any Author Accepted Manuscript version arising.

\appendix

\section{Amplitude Distribution for LIGO Livingston}
\label{app:L1amp}

\begin{figure}[h]
    \centering
    \includegraphics{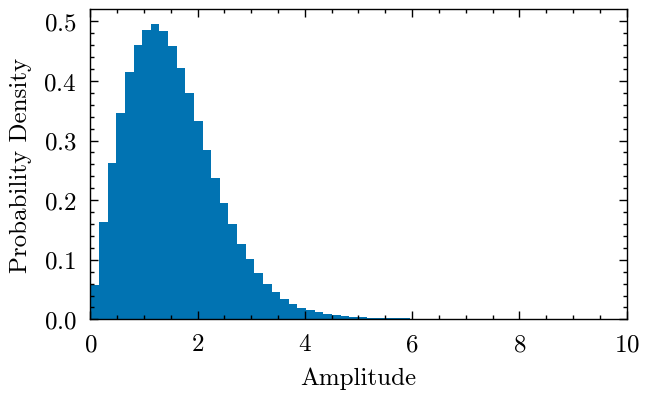}
    \caption{Distribution of amplitudes drawn for LIGO Livingston in the Monte-Carlo simulations.}
    \label{fig:L1amp}
\end{figure}

In Section \ref{ssec:Distance} we outlined incorporating a distance measurement into the amplitude terms present in the Monte-Carlo simulations. Here a maximum distance was chosen to normalize the amplitude distributions to a scale that is similar to a physical signal-to-noise ratio. The resulting distribution for LIGO Livingston is shown in Figure \ref{fig:L1amp} corresponding to a $D_{max}$ of 0.6. This $D_{max}$ is then scaled by the lowest relative sensitivity among detectors in the network, this ensures an appropriate number of samples are kept after thresholding. The relative sensitivities used in this study followed those used in the PyCBC configuration for the second part of the fourth observing run (O4b). These are 1.0, 0.94 and 0.32 for LIGO Livingston, LIGO Hanford and Virgo respectively.

\section{Flow performance in low-density regions} \label{app:density comparison}

\begin{figure}[H]
    \centering
    \includegraphics{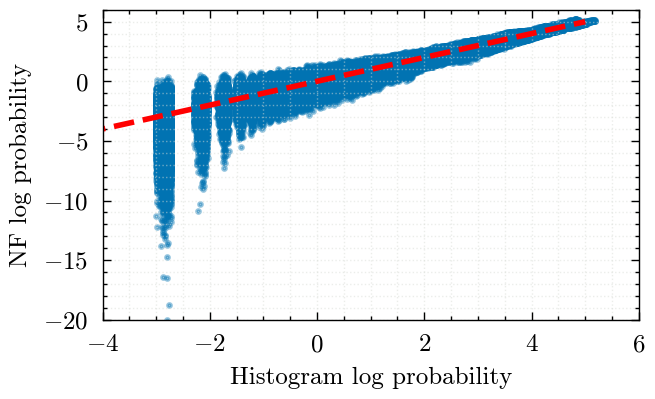}
    \caption{Comparison of the density stored in each HL histogram bin with the density predicted by the normalizing flow at the corresponding location. The red line corresponds to equal densities.}
    \label{fig:low density}
\end{figure}

The results shown in Figures \ref{fig:2DET_found_injs} and \ref{fig:3DET_found_injs} show that sensitivity is preserved using the normalizing flow. However, these searches are most sensitive to the high density regions of the parameter space where the bulk of signal lie. Here we look at how the density estimated using the two methodologies compares across the whole parameter space. In Figure \ref{fig:low density} we show the density stored in each histogram bin from the file used for the HL 'Modified' search with the density predicted by the normalizing flow at the corresponding location. We can see the uncertainty between the two methodologies increases towards the lower densities and that in the lowest weighted bins there is a more discrete distribution of densities for the histogram-based approach. Whereas, the flow models a continuous distribution in these spaces. This is due to the lower density regions being difficult to represent accurately with the histogram-based approach as many of the lowest density bins receive no samples and are populated only through a smoothing function. However, these regions correspond to combinations of time delays, phase delays and amplitude ratio that are either unphysical or have extremely low signal probability. Their role in the ranking statistic is primarily to down-weight noise triggers that exist in these unphysical regions rather than to identify astrophysical signals.

\section{Training a Normalizing Flow on Four and Five Detector Cases} \label{app:training parameter 4+}

\begin{figure*}[p] 
    \centering
    \includegraphics[width=\textwidth]{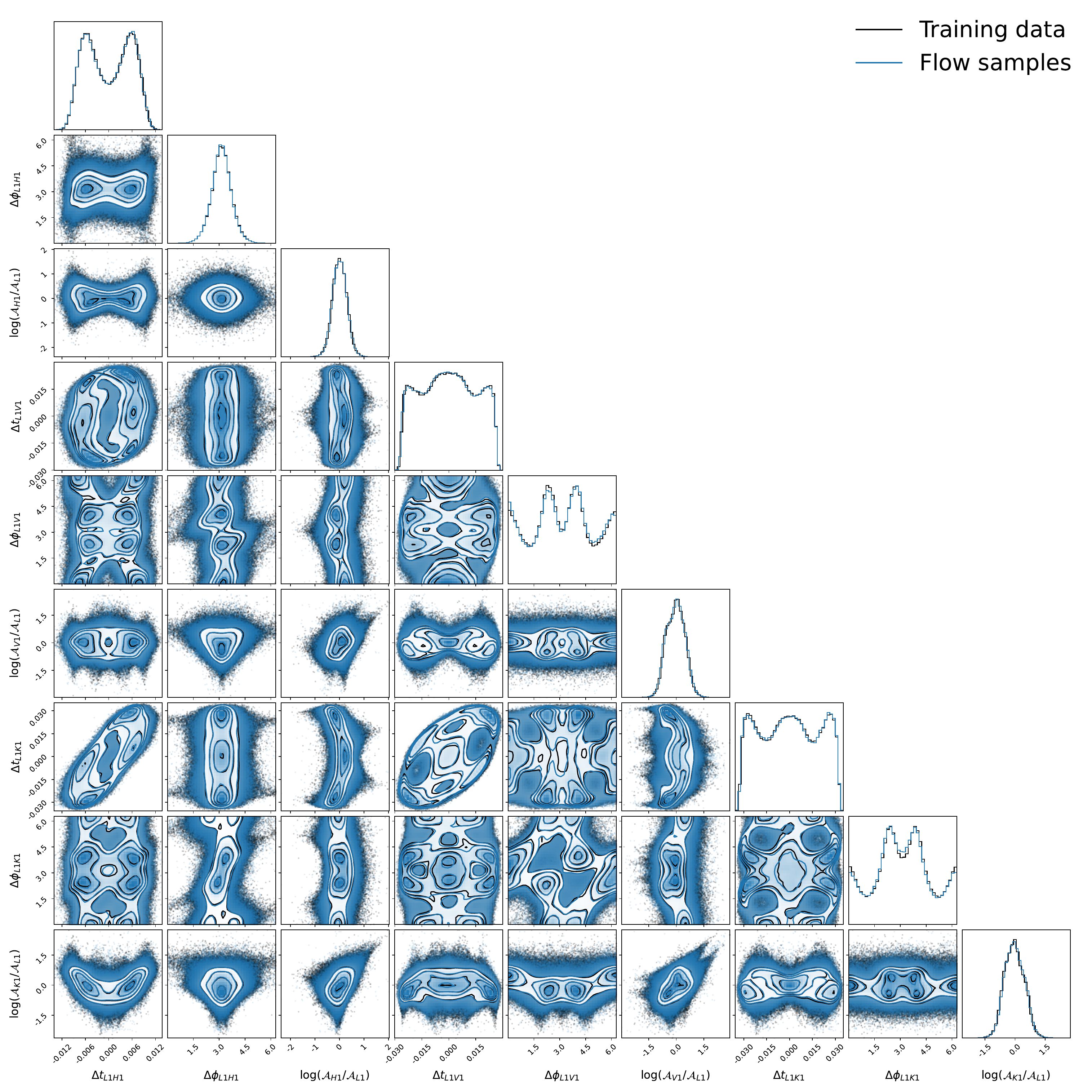}
    \caption{Corner plot comparing the sample distribution (black) and the learned distribution (blue) from a normalizing flow for a four detector network. Four equally sensitive detectors were used at the locations of LIGO Hanford (H1), LIGO Livingston (L1), Virgo (V1) and KAGRA (K1). The samples used were generated through the modified sampling methodology outlined in this paper. The normalizing flow was trained on the parameters shown in Table \ref{tab:flow training parameters 4+}.}
    \label{fig:4corner}
\end{figure*}

\begin{figure*}[p]
    \centering
    \includegraphics[width=\textwidth]{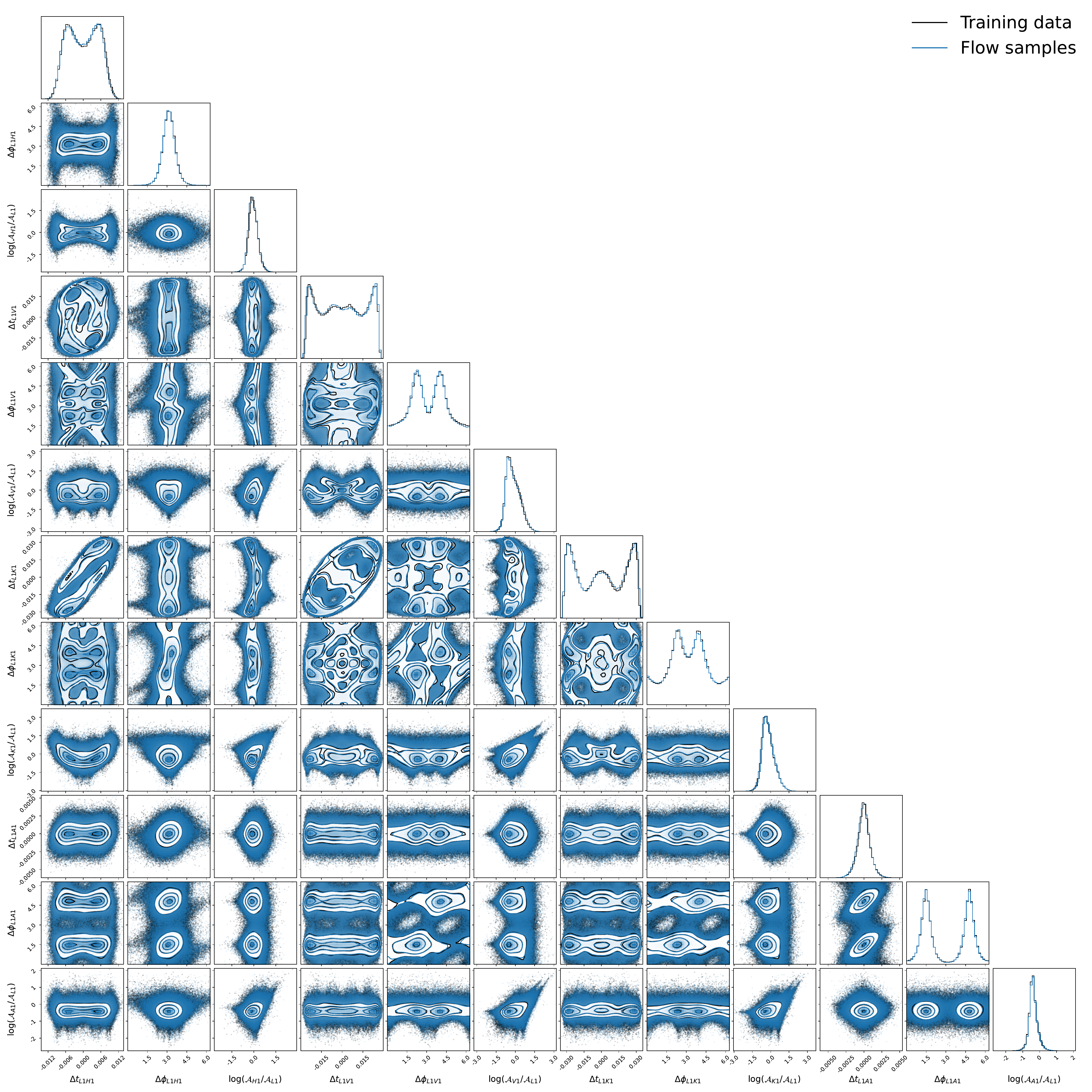}
    \caption{Corner plot comparing the sample distribution (black) and the learned distribution (blue) from a normalizing flow for a five detector network. Five equally sensitive detectors were used at the locations of LIGO Hanford (H1), LIGO Livingston (L1), Virgo (V1), KAGRA (K1) and LIGO India (A1). The samples used were generated through the modified sampling methodology outlined in this paper. The normalizing flow was trained on the parameters shown in Table \ref{tab:flow training parameters 4+}.}
    \label{fig:5corner}
\end{figure*}

For both four and five detector distributions, the normalizing flow was trained following the same process as outlined in Section \ref{ssec:Flow implementation}. Here, the training parameters were increased to account for the higher dimensionality of the distributions (9 dimensions for four detectors, 12 dimensions for five detectors), and are shown in Table \ref{tab:flow training parameters 4+}. We were not able to generate the corresponding files for the histogram-based approach due to extreme storage requirements. Hence, we show a comparison of the modeled distribution to the distribution of training samples, which are shown in Figures \ref{fig:4corner} and \ref{fig:5corner}. The modeled distribution is obtained by generating samples from the trained normalizing flow. This is done through the transformation of samples drawn from the latent distribution using the learned bijective mapping. In both cases we find a good match between the two distributions.

\begin{table}[H]
\centering
\begin{tabular}{|c|c|c|}
\hline
\textbf{Training Parameters} & \textbf{4 Detector}  & \textbf{5 Detector} \\ 
\hline
Number of transforms & 4  & 4 \\
Number of neurons &  128  & 140 \\
Latent distribution & Uniform  & Uniform \\
Number of bins & 20 & 25 \\
Number of training samples & 700,000 & 1,000,000 \\
Training Cost (CPU hours) & 3.25 & 8 \\
\hline
\end{tabular}
\caption{Example parameters used to train the normalizing flow in a four and five detector case. A comparison of the trained models against the sample distributions are shown in Figures \ref{fig:4corner} and \ref{fig:5corner}. }
\label{tab:flow training parameters 4+}
\end{table}

\clearpage
\bibliography{references}
\end{document}